\documentclass{llncs}
\usepackage[utf8]{inputenc}
\usepackage{authblk}
\usepackage{graphicx}
\usepackage{color}

\newcommand{\lqu}{~\lq\lq}
\newcommand{\rqu}{\rq\rq}

\title{On RAF sets and autocatalytic cycles in random reaction networks}
\author{Alessandro Filisetti\inst{1} \and Marco Villani\inst{2,1} \and Chiara Damiani\inst{3} \and Alex Graudenzi\inst{3} \and Andrea Roli\inst{4} \and Wim Hordijk\inst{5} \and Roberto Serra\inst{2,1}}

\institute{European Centre for Living Technology, University Ca' Foscari of Venice, Italy
\and University of Modena and Reggio Emilia, Department of Physics, Informatics and Mathematics
\and University of Milano Bicocca, Department of Informatics, Systems and Communication 
\and University of Bologna, Dept. of Computer Science and Engineering (DISI)
\and SmartAnalytiX.com, Lausanne, Switzerland}

\begin{document}
\maketitle

\begin{abstract}
The emergence of autocatalytic sets of molecules seems to have played an important role in the origin of life context. Although the possibility to reproduce this emergence in laboratory has received considerable attention, this is still far from being achieved.\\
In order to unravel some key properties enabling the emergence of structures potentially able to sustain their own existence and growth, in this work we investigate the probability to observe them in ensembles of random catalytic reaction networks characterized by different structural properties.\\
From the point of view of network topology, an autocatalytic set have been defined either in term of strongly connected components (SCCs) or as reflexively autocatalytic and food-generated sets (RAFs).\\
We observe that the average level of catalysis differently affects the probability to observe a SCC or a RAF, highlighting the existence of a region where the former can be observed, whereas the latter cannot. This parameter also affects the composition of the RAF, which can be further characterized into linear structures, autocatalysis or SCCs.\\
Interestingly, we show that the different network topology (uniform as opposed to power-law catalysis systems) does not have a significantly divergent impact on SCCs and RAFs appearance, whereas the proportion between cleavages and condensations seems instead to play a role.\\
A major factor that limits the probability of RAF appearance and that may explain some of the difficulties encountered in laboratory seems to be the presence of molecules which can accumulate without being substrate or catalyst of any reaction. 
\end{abstract}

\section{Introduction}
The dynamics of sets of interacting molecular species, in those cases where new molecular types can be created by the interactions among some of the existing ones, pose formidable problems. This study is of the outmost importance in researches on the origin of life (OoL) but its applications might be interesting also within laboratories or industrial applications.

When only few new molecular types can be generated, the complications are manageable. The actual interesting case is the one where many types might appear. In order to study the generic features of this kind of systems, the use of random molecular species has been introduced~\cite{Filisetti:2010mz,Filisetti2011a,Filisetti:2010fk,Filisetti2011b}, in particular in the context of the so-called binary polymer model~\cite{Kauffman:1986mi,Farmer1986,Kauffman:1993la} where all the species are linear strings made out of a binary alphabet. Since catalysis plays a prominent role in biological processes, it is often also assumed that only catalyzed reactions take place at an appreciable rate. In the spirit of searching for generic properties, randomness is also extensively used to determine catalysts: every molecular species has a certain fixed probability to catalyze a certain reaction.

In this way it is possible to represent the reactions as a directed graph, where there is an edge from a molecular species to all those species whose production reactions are catalyzed by that species. To complete the picture, it is worth stressing that only two types of reactions are allowed, condensation (which creates a new string by concatenating two previous ones) and cleavage (which generates two strings by separating two parts of a preexisting one). The model is described in more details in section~\ref{sec:model}.

The graph described above links catalysts to the products of the reactions they catalyze; it can therefore be considered as a \textit{catalyst $\rightarrow$ product} graph. The products may in turn catalyze other reactions. It has been often suggested that the presence of cycles (similar to those that are found in present-day biological systems~\cite{zlu:alberts02molecular,dyson-book,eigen-schuster1977,eigen-schuster-book,Farmer1986,jain-PNAS2001,Filisetti2011a,Filisetti:2010fk,segre1998}) plays a key role in the dynamics of these systems, since cycles of catalysts give rise to a collectively autocatalytic system. To be precise, we will call \textit{SCC} a Strongly Connected Component~\cite{diestel-graph-theory} of the network described above, i.e. a subset of the entire network where each node is reachable (directly or indirectly) from every other node in the subset. A \textit{SCC} is therefore composed of one or more molecular types where the formation of each member is catalyzed by at least another member of the \textit{SCC}.

While some models (e.g.~\cite{Jain:1998fk}) suppose that the presence of a catalyst is sufficient for a reaction to occur, in the binary polymer model considered here, in order for cleavage or condensation reactions to occur, the substrates must be present as well. These substrates may belong to what is called the food set F, which involves for example those molecules of the feed those are supposed to be freely available or to be continuously supplied from outside; these molecules might or might not have catalytic properties. Substrates
that are not in the food set may be present too, i.e. molecular species that take part in reactions and are produced and consumed by the system itself.


The importance of the substrates is properly captured by the notion of Reflexive Autocatalytic and Food generated (RAF) sets, that is a set of molecular types and reactions where each member can be generated by other members of the set through a series of catalyzed reactions, starting from $F$~\cite{Hordijk2011,Hordijk2012,Hordijk:2004vl}. Note that the identification of this structure needs knowledge about all the chemical species (substrates, products, catalysts) involved in the reaction scheme: a more detailed description can be found in section~\ref{sec:model}.

SCCs and RAFs are indeed important tools to analyze and understand the structure of sets of interacting molecules: note that an SCC may not be a RAF, given that substrates are not taken into account, and that a RAF may also not be an SCC, given that cyclic structures are not strictly required (a RAF may turn out not to have cycles if at least one of the species of the food set is also a catalyst).

Here below we analyze how frequent SCCs and RAFs are as a function of the average level of catalysis $\langle c \rangle$ (the average number of reactions catalyzed by each chemical species). Note that the behavior of the model turns out to be trivial when $\langle c \rangle$ is very small (no SCCs and no RAFs are observed) or very high (the system is full of both). Instead, it is interesting to consider an intermediate region. An outcome of the present work is that there exists a region of $\langle c \rangle$, similar results have observed in~\cite{Hordijk:2004vl} observing both RA and RAF sets, 
 that we term the gap region, where SCCs can be observed, but RAFs are not (these sentences are to be interpreted in a statistical sense, see the results in section~\ref{sec:res} for the details). Therefore, in this region, the very existence of cycles in the catalyst $\rightarrow$ product networks does not imply that they actually affect in a sensible way the main network properties.

In the present work we will investigate protocell systems, compartments containing sets of molecules able to collectively self-replicate, able to undergo fission and proliferate~\cite{Szostak:2001xw,Rasmussen2003}. Several protocell architectures have been proposed~\cite{Serra:2006aa,Carletti:2008rm,Filisetti2010b,Ganti:2003aa,Luisi2006,Mansy:2008kx,Rocheleau:2007gf,Sole:2007jw,Stadler1991}, most of them identify the compartment with a lipid vesicle that may spontaneously fission under suitable circumstances.  In the specific, we are here interested on a very simple protocell model, where a semi-permeable membrane embrace a small physical space where all the key reactions take place in the aqueous phase, and it allows the passage of small molecules only~\cite{Serra2014}.

In the present work we investigate sets of autocatalytic reactions where a constant concentration of some small chemical species is guaranteed: this situation can be maintained by chemo-physical systems as protocells (compartments containing sets of molecules able to collectively self-replicate, able to undergo fission and proliferate~\cite{Szostak:2001xw,Rasmussen2003,Serra2014}---an interesting system for both Origin of Life and technological themes) or CSTRs (Continuous-flow Stirred-Tank Reactor~\cite{schmidt-book}).

Thus, the topological structures having the highest probabilities to emerge in such systems, by taking a purely graph theoretical approach, will be investigated, as in previous papers by Kauffman, Steel and Hordijk~\cite{Hordijk:2004vl,Hordijk2012}, and the existence of the gap region will be shown. It is however tempting to speculate about the behavior of a truly dynamical model in this region.

We have already shown in our previous works~\cite{Filisetti:2010mz,Filisetti:2010fk,Filisetti2011b,Filisetti2011a} that, for $\langle c \rangle$ values close to the threshold for SCCs appearance, those that are observed tend to be fragile, some of their links refer to very rare reactions and do not really affect the behavior of the system~\cite{Filisetti:2010fk,Filisetti2011a}. We guess that this behavior is related to the absence of RAFs, and that by exploring the region close to the second threshold (i.e. the one for RAFs) one should be able to find more productive networks.
Moreover, we noted that (within the size of the systems we explored) the thresholds of SCCs and RAFs appearance in uniform and power-law catalysis systems are very similar. Interestingly, RAF systems at different average levels of catalysis are composed of different mixtures of autocatalysis, SCCs and linear structures. Different proportions of cleavages and condensations lead to different probabilities of RAFs appearance, and the presence of a large fraction of reactions producing long molecules (condensations ligating long chemical species) could inhibit the RAFs appearance as we will be discussed in section~\ref{sec:res}. So, the RAF presence is influenced in different ways by different characteristics of the chemical system.

In section~\ref{sec:res} the main results are presented, while the discussion of further work is deferred to the final section~\ref{sec:end}.

\section{Description of the model and network topologies}
\label{sec:model}

A detailed description of the theoretical model can be found in Kauffman~\cite{Kauffman:1993la,Kauffman:1986mi} and in Filisetti et al.~\cite{Filisetti:2010fk,Filisetti2011a}. In the following only the main features, useful for the scope of this work, will be summarized.
Random catalytic reaction networks are composed of a set of species $S=(s_{1}, s_{2}, \cdots, s_{n})$. In particular, species are concatenation of characters from left to right taken from a finite alphabet, in this work a binary alphabet $A=\{1;0\}$. Thus, the cardinality $|S|$ of $S$ is $|S|=2^{M+1}-2$, where $M$ is the maximum length of the system.
In accordance with the original version of the model \cite{Kauffman:1986mi}, each $s_{i} \in S$ can take part to two types of reaction only: condensation, where two species are concatenated to create a longer species (e.g. $100+10\rightarrow10010$), and cleavage, where a species is divided in two shorter species (e.g. $1001\rightarrow10+01$); $s_{i}$ can be either a substrate or a product according to the position within the reaction scheme.
Each species has an independent probability $p_{s_i}=1/|S|$ to be selected as a substrate for a reaction, hence in case of condensation two species will be selected while in case of cleavage just one species will be selected.
The overall number of conceivable reactions $\hat{R}$, i.e., all the reaction schemes allowed by the combination of all the possible $|S|$ molecular species, varies according to the constructional method adopted; in this work we typically adopt the method described in the following.\footnote{The only exception is presented in the final part of the results section.}

Reactions are created so that the product of the reaction cannot be longer than $M$, hence the number of conceivable reactions is equal to:
\begin{equation}
\hat{R}=2 \cdot \sum_{i=2}^{M}2^i \cdot (i-1)
\label{eq:conceivableRcts2}
\end{equation}
where the multiplicative term $2$ indicates that each reaction scheme is in principle valid both for cleavage reactions and condensation reactions.\footnote{Forward and backward reactions, during the creation of the reactions graph, are in principle handled as two separated reactions.}~
Not all the reactions belonging to $\hat{R}$ are catalyzed. So, it is possible to compute an average level of catalysis $\langle c \rangle = R/|S|$ which denotes the average number of reactions catalyzed by each single species.\footnote{Since different $M$ will be assessed, in this work we prefer to adopt the average level of catalysis $\langle c \rangle$, as in~\cite{Hordijk2011}, instead of the standard reaction probability $p$, as we adopted in our previous works~\cite{Filisetti2011a,Filisetti:2010fk}, and Kauffman ~\cite{Kauffman:1986mi} and others~\cite{R.J.Bagley:1991ys,Mossel2005} in important works on this topic. Nevertheless, according to $|S|$, it is always possible to move from $\langle c \rangle$ to $p$ and \textit{vice-versa}. }
Therefore, each catalysis can be represented as a pair $(r_{k},s_{i})$ where $r_{k}\in\hat{R}$ stands for the selected reaction and $s_{i}\in S$ behaves as a catalyst for $r_{k}$.

The procedure of catalysis assignation leads to different catalytic reaction network topologies. 
In this work, two kinds of topologies will be assessed, based on uniform and preferential assignation. In the former case, each $s_i \in S$ has the same probability $p^i_k=1/|S|$ to be a catalyst for a whatever reaction $k$, while in the latter case $p^i_k$ is determined by the chemical characteristics of the involved species. In real systems, researchers observed that very few species can catalyze many reactions, whereas many species catalyze only one reaction, or none. This situation can be represented by means of a catalysis distribution having the shape of a power law, our second topology.\footnote{We obtain the power law distribution by slightly modifying the algorithm proposed by Barab\'asi and Albert~\cite{barabasi-reka-science1999}. We increase $p^i_k$ as a function of the already catalyzed reactions, i.e., the probability to catalyze a reaction is weighted with the number of reactions already catalyzed so that $p^i_k=\#r_i/\sum_{z=1}^{|S|} \#r_z$, where $\#r_i$ and $\#r_z$ indicate the number of reactions already catalyzed by the $i$-th and the $z$-th species respectively. In such a way, we obtain a power-law distribution in the number of reactions each chemical species can catalyze.}

\begin{figure}[ht]
\begin{center}
\includegraphics[width=12cm]{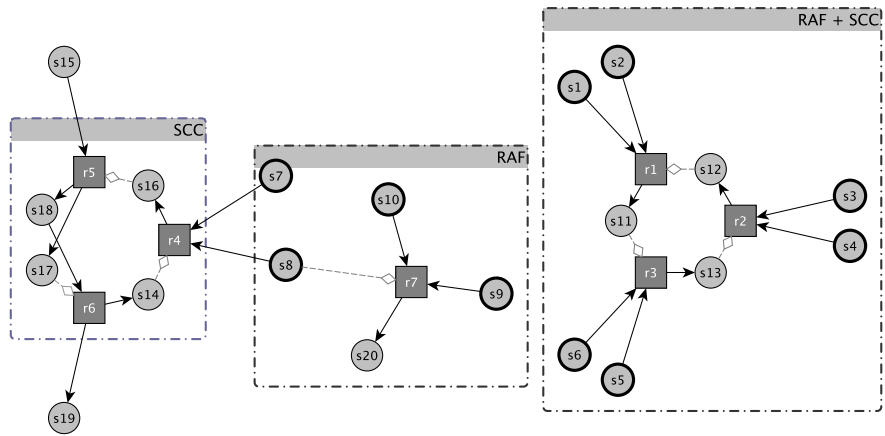}\\
\caption{The picture shows an example of representation of a catalytic reaction network by means of a complete reaction graph, taking into account both molecular species and reactions. Circles stand for molecular species and in particular bold circles represent species belonging to the food set $F$. Squares depict reactions, straight arrows indicate the participation of a species to the reaction as substrate (edge points to the reaction) or product (edge starts from the reaction). Catalysis are represented by dotted gray arrows. For the sake of comprehension, both molecular species and reactions forming RAFs and SCCs are grouped together.}
\label{fig:ccrn}
\end{center}
\end{figure}

\begin{figure}[ht]
\begin{center}
\includegraphics[width=4cm]{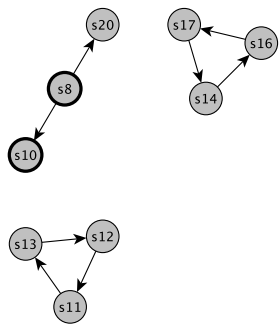}\\
\caption{The picture shows the catalyst $\rightarrow$ product representation of the complete graph $\mathcal{C}$ shown in figure~\ref{fig:cpgraph}. In this representation just the catalytic activity of the reaction network is depicted, and no information about the substrate of the reactions are present. Circles stand for molecular species and in particular bold circles represent species belonging to the food set $F$. In this case straight arrows indicate the catalysis of a particular molecular species by means of its catalyst. By means of this representation, the two SCCs are clearly visible and they are formed from $s_11, s_12, s_13$ and $s_14, s_16, s_17$ respectively.}
\label{fig:cpgraph}
\end{center}
\end{figure}

The entire set $\mathcal{C}=(r_{k},s_{j})$ of catalysis form what we call~\lq\lq artificial chemistry\rq\rq, hence different random instantiations of $\mathcal{C}$ lead to different random catalytic reaction networks, i.e., different~\lq\lq artificial chemistries\rq\rq; in figure~\ref{fig:ccrn} an example of a complete reaction graph is depicted while in figure~\ref{fig:cpgraph} the catalyst $\rightarrow$ product representation of the graph shown in figure~\ref{fig:ccrn} is reported.

The structural role of backward reactions is evaluated as well. Thus, artificial chemistries imposing the presence of both forward and reverse reactions will be created.\footnote{Since the analysis are static and only statistic structural properties of the networks will be assessed, the choice on which reaction is the direct one and which is the reverse one is only implementative and does not affect the analysis.} It is worthwhile to notice that introducing reverse reactions leads to a double number of reactions catalyzed, so in case of reverse reactions the number of reactions to catalyze is divided by $2$, $\hat{R} \rightarrow \hat{R}/2$.

Analysis will be carried out measuring the emergence of reflexively autocatalytic sets (RAF sets) on the complete graph $CRN=(S,R,\mathcal{C})$, where $\mathcal{C}$ represents catalysis,  and investigating the strongly connected components emerging in the graph representation considering catalysts and products only, hence showing the catalytic activity of the system without taking care of the presence and the nature of the substrates. In figure~\ref{fig:ccrn} two examples of RAF sets and two SCCs are shown. In particular, the group of molecular species and reactions belonging to the group named~\lq\lq $RAF + SCC$\rq\rq~ form a SCC and a RAF at the same time. On the opposite, the other two reaction structures are, from left to right, a SCC only and a RAF set only.
\section{Results}
\label{sec:res}

As discussed above, we investigate here catalytic reaction networks, where only small molecular species have a constant concentration (this situation can be maintained by protocells~\cite{Serra2014}, or CSTR systems~\cite{Filisetti:2010fk}).
In the following, scenarios where the length of the buffered species (the~\lqu foodset\rqu~of the from which RAF structures are built) ranges from $2$ ($|S|=6$) to $4$ ($|S|=62$) will be considered.

We are interested now in analyzing the role of different scenarios on both the presence and the internal structure of RAFs within protocells. The assessed scenarios involve different distributions of the chemicals' catalyzing capabilities (a uniform distribution, and a slightly more realistic power-law distribution, which allows the presence of a little number of highly versatile catalysts), the presence or absence of backward reactions, and the possibility for long chemical species to be not suitable for participating in any reaction, leading in such a way to the accumulation of useless chemical species (for example, because lack of solubility or folding difficulties).

\subsection{Strongly Connected Components and RAFs}
\label{sccraf}

\begin{figure}[ht!]
\begin{center}
\includegraphics[width=5cm]{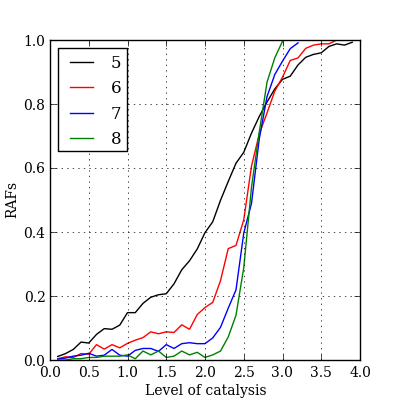}
\includegraphics[width=5cm]{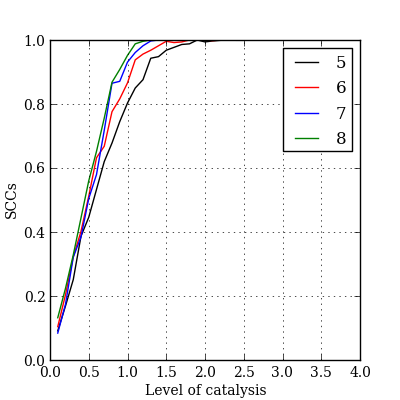}
\caption{The fraction of simulations showing at least $1$ RAF (left) and $1$ SCC (right), by varying $\langle c \rangle$ and $M$ in networks with forward reactions only. $F$ contains all the species up to length $2$. On the x-axis the average level of catalysis $\langle c \rangle$ is represented while on the y-axis the fraction of network instances (out of $1000$ networks for each $\langle c \rangle$) is depicted.}
\label{fig:ACSfraction_food2}
\end{center}
\end{figure}

The first observation regards the presence of SCCs and RAFs within the reaction graphs in accordance with a foodset composed of species up to length $2$ only, i.e., when the membrane allows the transfer of chemicals having maximum length $2$. Figure~\ref{fig:ACSfraction_food2} shows the fraction of network instances showing at least $1$ SCC ($1$ RAF), by varying $\langle c \rangle$ and $M$. It can be observed that the SCCs and RAFs transition zones significantly differ (the higher the maximum length of the system, the higher the transition slope): the vast region between $1.0$ and $2.5$ levels of catalysis is therefore rich of SCC structures unable to self-sustain because the failed fulfilling the closure condition.~\footnote{It is important to notice that in~\cite{Hordijk:2004vl} the transition is found at $1.25$. Such a difference is basically related to a different way to count forward and reverse reactions. Hordijk~\cite{Hordijk:2004vl} consider both forward and backward reactions as a unique reaction whereas in our work, though it is clear that they account for the same reaction scheme, we consider them as two different reactions.}
 
The showed case regards scale-free topologies without backward reactions (obviously, in these experiments we maintain constant the total number of reactions). It is worth mentioning that  the bias induced on the system topology by the pairing of forward-backward reactions do not change the positions of the transition zones nor their slopes (data not shown here), as long as condensation and cleavages are equally present - we will discuss the effect of uneven presence of condensation and cleavages in the next paragraph. Remarkably, almost identical behaviors hold for random topologies having the same number of chemical species and reactions (data not shown, and~\cite{Hordijk:2004vl}).\footnote{In~\cite{Hordijk:2004vl}, the authors show the so-called RA sets---which are reflexively autocatalytic but not necessarily food-generated---whose transition and nature correspond to that of SCC, where the transitions happen on the same zones of scale-free topologies in case of non catalytic activity of the foodset molecular species, i.e,~\lqu $BUF_2$\rqu~scenario in this work.}

 
\begin{figure}[ht!]
\begin{center}
\includegraphics[width=4cm]{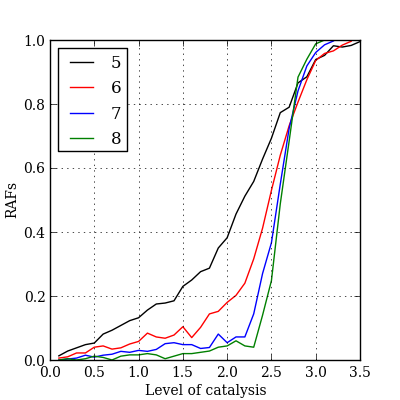}
\includegraphics[width=4cm]{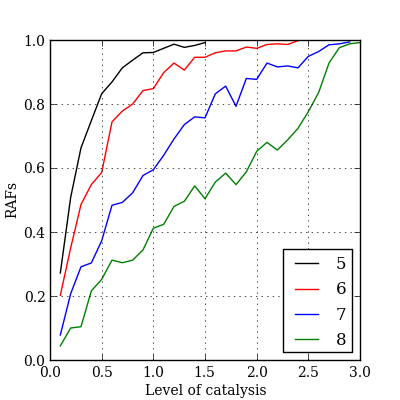}\\
\includegraphics[width=4cm]{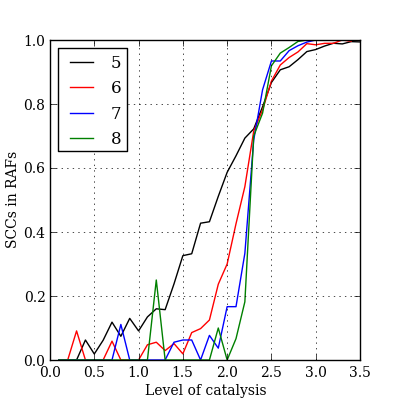}
\includegraphics[width=4cm]{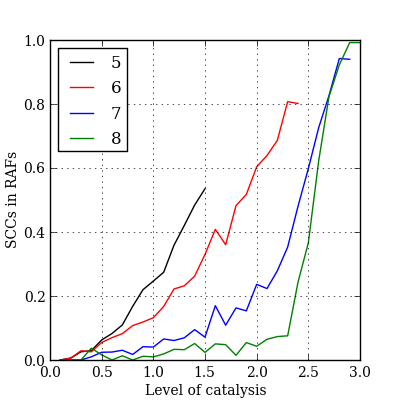}\\
\includegraphics[width=4cm]{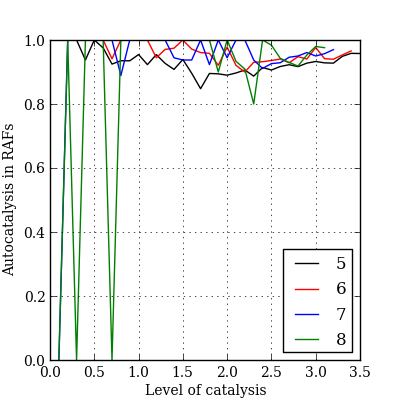}
\includegraphics[width=4cm]{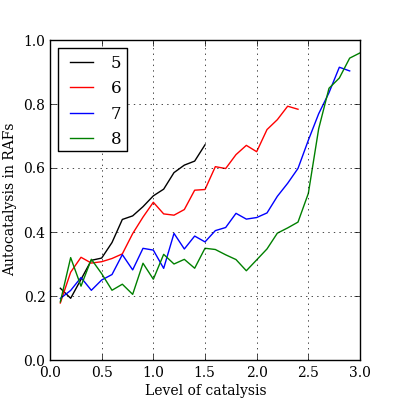}
\caption{TOP:  Fraction of simulations showing at least $1$ RAF.  MIDDLE:  fraction of RAFs having at least $1$ SCC. BOTTOM: fraction of RAFs having at least $1$ autocatalysis. On the x-axis $\langle c \rangle$ is represented, $F$ is composed of all the species up to length 2 (left panel) and 3 (right panel). For each $\langle c \rangle$ and for each value of $M$, $1000$ network instances have been created. For computational reasons, once the $100\%$ of networks with a specific $\langle c \rangle$ contain at least a RAF set, the system automatically goes to the next $M$, thus in some cases the analysis on SCC does not reach a $\langle c \rangle=4$, the maximum level of catalysis evaluated.}
\label{fig:ACSfraction_food3}
\end{center}
\end{figure}
\subsection{The inner RAF structure}
\label{sec:innraf}

The aforementioned results are derived from systems with protocell membranes that allow the diffusion of chemical species up to length $2$ (\lqu $BUF_2$\rqu~scenario). Since $F$ does not show any catalytic activity, in that case all RAFs are necessarily composed of at least one SCC.  On the contrary, if we extend the membrane semi-permeability to longer species, some of these objects may be able to catalyze some chemical reactions: this is a crucial change, because the formation of RAFs composed of linear structures (on the catalyst $\rightarrow$ products graph) turn out to be possible, thus catalytic roots are buffered (\lqu $BUF_3$\rqu~ scenario).

This change in $F$ composition has an apparently huge effect on the presence of RAF, which appears also at low level of catalysis (figure~\ref{fig:ACSfraction_food3}); note however that the difference among chemistries with different maximum lengths $M$ (\lqu MaxLen $M$\rqu~systems) consists mainly in lower and lower probabilities of choosing the buffered species as root of linear structures on the catalyst $\rightarrow$ products graph (a more interesting use of the \lqu MaxLen $M$\rqu~scenarios is presented in the next paragraph). Indeed, the presence of the $2.5$ threshold can be grasped on the final part of~\lqu MaxLen $8$\rqu~scenarios: it is a clue that the line showing greatest chemistries (where there is a very low probability of choosing small molecules as catalysts) tends towards the plot of~\lqu $BUF_2$\rqu~scenario (where small molecules are not catalysts). Such ample systems (and the~\lqu $BUF_2$\rqu~scenario as well) may represent a good approximation of real chemistries.

More interesting differences are observed when comparing the fraction of systems having at least one SCC or one autocatalysis within the obtained RAFs.~\footnote{It is worth stressing that an autocatalysis is a strongly connected component, nevertheless we decided to deal with them separately.}
At low catalysis levels, within the~\lqu $BUF_2$\rqu~scenario, almost all RAFs contain an autocatalysis, whereas the formation of other SCC structures inside the RAF is unlikely; on the contrary, as the average catalysis level grows up, the fraction of RAFs entailing an SSC tends to $1$ (slightly before the $2.5$ zone) - note that these RAFs still have an high probability of containing an autocatalysis as well. The situation within the~\lqu $BUF_3$\rqu~scenario differs substantially: the sum of SCCs and autocatalysis do not reach the $100\%$, and this gap increases as the maximum allowed length decreases, that is, the prevailing structures for a large zone of catalysis level are linear chains on the catalyst $\rightarrow$ product graph. SCCs and autocatalysis play a minor role.~\footnote{At high catalysis levels almost each RAF owns at least one SCC and one autocatalysis, so with our measurements it is not possible to observe the exact RAFs' structure. This aspect will be analyzed in further works.}

Within a scenario that allows the diffusion of chemical species up to length 4 (data not shown here), the growth in the probability to observe a RAF happens even earlier than in the \lqu $BUF_3$\rqu~scenario, whereas the presence of SCCs and autocatalysis within the RAF is almost identical between the two cases (indeed, the probability of having SCCs and autocatalysis are identical, not depending on the number of the buffered chemical species). So, in~\lqu $BUF_4$\rqu~scenario the presence of linear chains on the catalyst $\rightarrow$ product graph is even more significant.  Remarkably, similar graphs and values hold for the scale-free topologies we tested (same number of chemical species and reactions).
\subsection{Cleavages, condensations and irreversibility}
\label{sec:irreversibility}

Chemical reactions are generally reversible, in the sense that the reaction proceeds both forward (from reactants to products) and backward (in the opposite direction); however, note that, when far from equilibrium, the rates of the two processes are not equal. Particular environments (as aqueous phases rather than oleic phases) can hugely influence the reaction rates of all cleavages and/or of all condensations, whereas the presence or absence of catalysts can hugely influence the reaction rates of some specific reactions (for instance the particular sets of compartments, chemical species and reactions present inside the current living systems can hugely influence the reaction rates inside cells).

Hence, in case of similar rates the proportion between cleavages and condensations is $50:50$, whereas in case of particular environments or particular living systems this proportion could constitute a relatively free parameter. In neutral environments the variation of this parameter do not have particular consequences: however, this is not the case where the concentration of small species is kept constant.

\begin{figure}[ht!]
\begin{center}
\includegraphics[width=4cm]{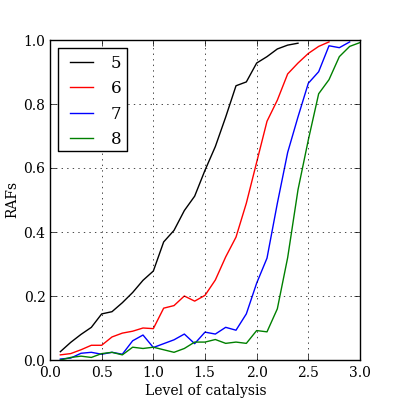}
\includegraphics[width=4cm]{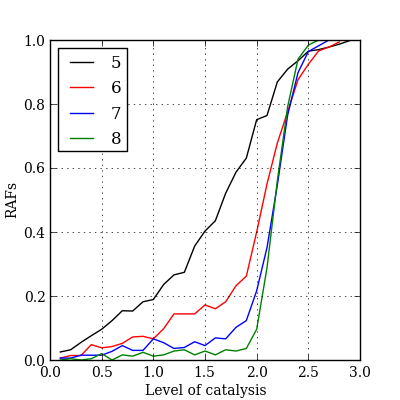}
\includegraphics[width=4cm]{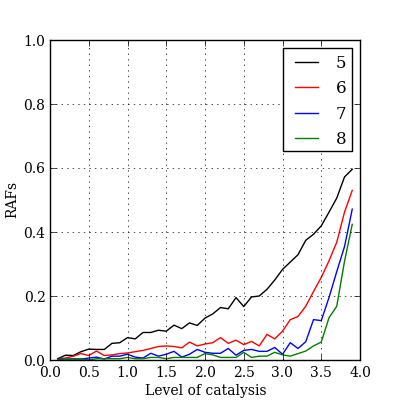}
\caption{The fraction of simulations with a food set composed of all the species up to length $2$ showing at least  $1$ RAF, by varying the average level of catalysis and the maximum chemical species length. The plots show situations where the proportion between condensations and cleavages is $100:0$ (left), all condensations; $75:25$ (middle); $25:75$ (right). The case $0:100$ (all cleavages) is devoid of RAFs, hence graph is not shown here while the case $50:50$ is shown in figure~\ref{fig:ACSfraction_food3} left top.}
\label{fig:differentrcc}
\end{center}
\end{figure}

Indeed, since the membrane tends to allow the free transfer to small molecules only, the apparent symmetry of this situation is broken: whereas it is possible to continuously construct long molecules by simply concatenating (by means of recursive condensation reactions) the always present small ones, it is not possible to continuously cleave long molecules, because their presence is not similarly guaranteed. Thus, it is not possible to observe RAF if just cleavages are present. Figure~\ref{fig:differentrcc} shows the influence of different levels of the proportion between cleavages and condensations on the presence of RAF in chemical sets and on the position of their previously commented transition, in the case~\lqu $BUF_2$\rqu~scenario. In this case, note that, because the product of reactions involving buffered molecules are buffered molecules and because these buffered molecules do not catalyze anything, in presence of only cleavages nothing happen (data not shown here).
 
\begin{figure}[ht!]
\begin{center}
\includegraphics[width=5cm]{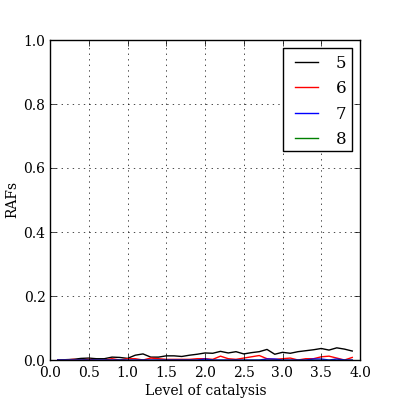}
\includegraphics[width=5cm]{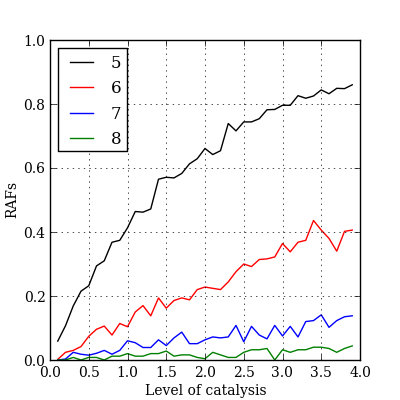}
\caption{Fraction of random systems, with the~\lqu purely combinatorial\rqu~quotient between cleavages and condensations, having at least $1$ RAF, with food till length 2 and (c) till length $3$}
\label{fig:m1}
\end{center}
\end{figure}
Note that if we allow all the possible condensations and cleavages among the existing molecules up to length $L$, the relative proportion of cleavages and condensation is obtained by a purely combinatorial argument. 
In this case the number of conceivable reactions is equal to:
\begin{equation}
\hat{R}_1=\sum_{i=2}^{L}[2^i \cdot (i-1) ]+|S|^{2}
\label{eq:conceivableRcts}
\end{equation}

where the first term stands for the cleavage reactions and the second term for the condensation reactions~\cite{Filisetti:2010fk}. It is worth stressing that, as $L$ increases and preventing the presence of reverse reactions, the number of conceivable condensation reactions tends to become higher than the number of conceivable cleavages, according to the ratio  $[\sum_{i=2}^{L}2^i \cdot (i-1)] / |S|^2$. In other words, the~\lqu purely combinatorial\rqu~proportion between cleavages and condensations is in favor of condensations, and this advantage increases with the diversity of the present chemical species.

Following this idea, there may exist lots of condensations whose products are longer than $L$.
This second method, applied to the same number of chemical species, leads to a very different number of reactions: so the question arises as how can we compare these two methods, and what can we say about their physical plausibility. We can observe that in (the relatively simple) pre-biotic environment very long molecules rarely appear, because of generic physical-chemical constraints. Thus, we think that the more correct model interpretation of the $L$ limit should be:~\lqu chemical species longer than $L$ are not suitable for further chemical processes, because of physical-chemical constraints\rqu~(for example, lack of solubility). A large fraction of the condensations leads therefore to not suitable products, which cannot play an active role within the system; thus the creation of this kind of~\lqu garbage\rqu~can inhibit the formation of RAFs (see figure~\ref{fig:m1}).
\section{Conclusion}
\label{sec:end}

The comprehension of the properties of sets of many interacting molecules poses formidable problems, and it is still a big challenge to obtain them in laboratory.

In this work we aim at revealing some properties of abstract realizations of catalytic reaction networks, with particular regard to the presence of structures potentially able to sustain their own existence and growth (the Reflexively Autocatalytic Food generated sets, briefly RAFs). Our investigations take into account different scenarios, involving different distributions of the chemicals' catalyzing capabilities, the presence or absence of backward reactions, and the presence  or absence of long chemical species that lead to the accumulation of useless chemical species.

We confirmed that there is an ample region where the systems have autocatalytic structures, which are nevertheless unable to self-sustain (RAFs are not present); interestingly, this region shows the same amplitude across different topologies and/or settings. At a sufficient level of catalysis, complete RAFs emerge, whose presence and composition (different proportion of strongly connected components, autocatalysis and linear chains) depend on some parameters (average level of catalysis, number of buffered chemical species, proportion between cleavages and condensations) and is independent on other characteristics (uniform or scale-free topology, presence or absence of backward reactions). Finally, we discussed two different ways to build artificial networks and their physical implications.

We think that these hints could be useful for the comprehension of the emergence of sets of molecules able to collectively self-replicate in OoL scenarios, and for the designing of new artificial protocells.

\section*{Acknowledgements}
The research leading to these results has received funding from the European UnionSeventh Framework Programme (FP7/2007-2013) under grant agreement n. 284625 and from~\lq\lq INSITE - The Innovation Society, Sustainability, and ICT\rq\rq~Pr.ref. 271574, under the 7th FWP - FET programme. The final publication is available at Springer via \texttt{http://link.springer.com/chapter/10.1007/978-3-319-12745-3\_10}.

\footnotesize
\bibliographystyle{abbrv}
\bibliography{biblio}

\end{document}